\documentclass[letterpaper, 10 pt, conference]{ieeeconf}  
\usepackage{multirow}
\usepackage{graphicx}
\usepackage{amsfonts}
\usepackage{amssymb}
\usepackage{amsmath}
\usepackage{algorithm}
\usepackage{algorithmic}

\usepackage{color}
\usepackage{tikz}
\usetikzlibrary{arrows.meta,positioning}

\usepackage[natbib=true,style=numeric,sorting=none, maxnames=6]{biblatex}
\DefineBibliographyStrings{english}{%
  bibliography = {References},
}

\IEEEoverridecommandlockouts                              

\title{\LARGE \bf
A Data-Driven Model Predictive Controller to manage epidemics: The case of SARS-CoV-2 in Mauritius}

\author{S. Z. Sayed Hassen, Ahmed Aboudonia and John Lygeros
\thanks{This work was supported as a part of NCCR Automation, a National Centre of Competence in Research, funded by the Swiss National Science Foundation (grant number 51NF40\_225155)}.
\thanks{S. Z. Sayed Hassen, A. Aboudonia, and J. Lygeros are with the Automatic Control Laboratory, Swiss Federal Institute of Technology
(ETH) Zurich, Physikstrasse 3, 8092 Zurich, Switzerland {\tt\small ssayed@ethz.ch}}%
}

\begin{document}

\maketitle
\thispagestyle{empty}
\pagestyle{empty}

\begin{abstract}
  This work investigates the benefits of implementing a systematic approach to social isolation policies during epidemics. We develop a mixed integer data-driven model predictive control (MPC) scheme based on an SIHRD model which is identified from available data. The case of the spread of the SARS-CoV-2 virus (also known as COVID-19) in Mauritius is used as a reference point with data obtained during the period December 2021 to May 2022. The isolation scheme is designed with the control decision variable taking a finite set of values corresponding to the desired level of isolation. The control input is further restricted to shifting between levels only after a minimum amount of time.  The simulation results validate our design, showing that the need for hospitalisation remains within the capacity of the health centres, with the number of deaths considerably reduced by raising the level of isolation for short periods of time with negligible social and economic impact. We also show that the introduction of additional isolation levels results in a smoother containment approach with a considerably reduced hospitalisation burden.
\end{abstract}

\section{Introduction}
Since the appearance of the Severe Acute Respiratory Syndrome CoronaVirus-2 (SARS-CoV-2) in Wuhan, China, in December 2019, a fair amount of work has been undertaken by researchers in the control community to model and mitigate its progression; see for e.g., \cite{Kucharski2020, Chowdhury2020, Moore2021}. The spread of the pandemic depends on numerous social and economic factors. Different metrics have been used to determine the effectiveness of approaches developed to counteract the propagation of COVID-19. One of the most popular ones is the effective reproduction number $R_t$ \cite{Haug2020}, which is the average number of secondary cases that arise per infectious case in a population.

Authorities worldwide have adopted a variety of approaches both in their timing and duration to curb the spread of the virus with varying levels of success. In \cite{Cascini2022}, the effectiveness of different social containment policies based on evidence gathered in five European countries are compared. A similar comparison is made for a few Asian countries together with the United States, United Kingdom and France in \cite{Chen2021}. Both works conclude that strict early restrictions result in better control of the outbreak. 

Mauritius has been affected by three clearly defined SARS-CoV-2 waves. The first wave of COVID-19 was well controlled with very few casualties but the number of cases surfacing in the second and third wave stretched the capacity of health care facilities. Although the virus was well understood by then and health centres had developed a coordinated approach to handling infected cases, the main problem had been one of government preparedness to handle the number of new cases especially with the appearance of mutated variants of the virus. 

Here, we focus on the problem of better managing the spread of the virus in Mauritius using a dynamic non-pharmaceutical intervention approach. We first identify an SIHRD (Susceptible-Infected-Hospitalised-Recovered-Dead) model from data of the most severe wave of the epidemic which began in December 2021 and ended in May 2022. We resort to Genetic Algorithms for this task due to their ability to search for global optima in complex multi-dimensional spaces \cite{Khandelwal2019, Jain2024}. We then propose a data-driven MPC scheme through which the isolation levels (social distancing and sanitary precautions) of the country are regulated so that the number of infected cases remain within manageable limits, in turn keeping the number of hospital admissions within set bounds and minimising the number of casualties. Given the nature and dynamics of the spread of COVID-19, the MPC approach provides a fitting framework to mitigate its impact. Indeed, the approach has been used in a few research works related to the spread COVID-19. In Italy, for example, MPC was used to propose a structured multiregional approach to coordinate decision-making between different autonomous regions \cite{Carli2020}. Similarly, the computational strength of MPC and its inherent capacity to cater to constraints is leveraged in \cite{Morato2020} to design a binary ``on-off'' control scheme in Brazil. The MPC scheme also showed its robustness to model uncertainty and inaccurate measurements (sub-notifications of infected cases) in \cite{Kohler2021} when applied to the case of Germany. 

The main contributions of this work are:
\begin{itemize}
    \item Identify an SIHRD model using genetic algorithms from the dataset for Mauritius,
    \item Develop a data-driven MPC using the identified model taking into account the specific constraints and objectives of the case study,
    \item Validate the controller design through simulation to show that hospitalisation requirements remain within bounds and reduce the number of casualties,
    \item Explore the effect of varying design hyper-parameters on the MPC performance and modifying the number of isolation levels adopted.
\end{itemize}

In Section~\ref{sec:Modelling}, we explain the mathematical model used to represent the dynamics of the evolution of COVID-19 virus. Section~\ref{sec:NSI} presents the Genetic Algorithm approach used to identify the parameters of the nonlinear model. In Section~\ref{sec:MPC}, we determine the objective function and formulate the constrained optimisation problem to determine the most appropriate control policy to be adopted by the authorities. Numerical results to validate our approach and controller design are presented in Section~\ref{sec:Results} before we conclude in Section~\ref{sec:Conclusion}. 

\section{Modelling} \label{sec:Modelling}
\begin{figure}[htb]
	\begin{center}
	   \scalebox{1}{	    \begin{tikzpicture}[thick,scale=0.8, every node/.style={scale=0.8}]
			\node[draw,thick,rectangle,rounded corners=0.25cm,minimum size=.8cm] (S) {Susceptible (S)};
			\node[draw,thick,rectangle,rounded corners=0.25cm,minimum size=.8cm, right = 1.2cm of S] (I) {Infected (I)};
			\node[draw,thick,rectangle,rounded corners=0.25cm,minimum size=.8cm, below = 1.2cm of S] (H) {Hospitalized (H)};
			\node[draw,thick,rectangle,rounded corners=0.25cm,minimum size=.8cm, below = 1.2cm of I] (R) {Recovered (R)};
			\node[draw,thick,rectangle,rounded corners=0.25cm,minimum size=.8cm, left = 1.2cm of H] (D) {Dead (D)};
			\draw[-Stealth,thick] (S) -- (I) node[midway,above] {$\beta$};
			\draw[-Stealth,thick] (I) -- (H) node[midway,above] {$\lambda$};
			\draw[-Stealth,thick] (H) -- (D) node[midway,above] {$\rho$};
			\draw[-Stealth,thick] (H) -- (R) node[midway,above] {$\gamma_B$};
   		  \draw[-Stealth,thick] (I) -- (R) node[midway,right] {$\gamma_A$};
		\end{tikzpicture} }
	   \caption{SIHRD epidemic model used for COVID-19. The nodes categorise the ``total'' number of individuals at a given time and the arrows represent the direction and rate of flow of individuals from one category to the next.}
	   \label{fig:SIHRD-model}
    \end{center}
\end{figure}
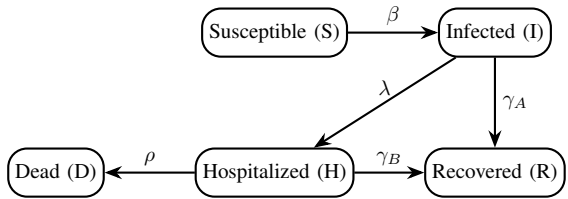

A mathematical representation of the evolution of epidemics using Susceptible-Infected-Recovered (SIR) model was first presented in \cite{Kermack1927}. Variations of that original model to include other ``states" have since been widely used, especially during the COVID-19 epidemic to model the evolution of the virus in Wuhan (China) \cite{Kucharski2020, Peng2020}. These states included the number of Exposed(E), Quarantined(Q) and Dead(D) among others. The data set available for the third wave of COVID-19 in Mauritius includes the number of Infected, Hospitalised and Fatalities. As such, we propose to use the SIHRD model where $S, I, H, R, \mathrm{and} \,D$ represent the number of Susceptible, Infected, Hospitalised, Recovered, and Dead individuals respectively. The corresponding epidemiological model used in this study is shown in Figure~\ref{fig:SIHRD-model}.

The model requires that fatalities only come about after hospitalisation and as such the model assumes that health care facilities always had the resources to admit sick patients. Furthermore, to model the dynamics of social distancing policies, we include an additional state $\Psi$ that reflect the response of the population to the isolation policies \cite{Morato2020}. $\Psi = 0$ stands for complete lockdown and $\Psi = 1$ corresponds to no isolation. The government policies are assumed to take on fixed values, between 0 and 1, with no isolation corresponding to $u=0$ and total isolation (complete lockdown) $u=1$. Given these definitions, and including the population dynamics into the SIHRD model, we can write the SIHRDC model as follows:
\begin{eqnarray}
  \frac{dS(t)}{dt} &=& - \Psi(t) \beta \frac{S(t)I(t)}{N}, \nonumber\\ 
  \frac{dI(t)}{dt} &=& \Psi(t) \beta \frac{S(t)I(t)}{N} - \gamma_A I(t) - \lambda I(t), \nonumber\\ 
  \frac{dH(t)}{dt} &=& \lambda I(t) - \frac{1}{1 - \rho} \gamma_B H(t), \label{eq:nonlinear-eqns} \\ 
  \frac{dR(t)}{dt} &=& \gamma_A I(t) + \gamma_B H(t),  \nonumber 
\end{eqnarray}
\begin{eqnarray}
   \frac{dD(t)}{dt} &=& \frac{\rho}{1-\rho} \gamma_B H(t), \nonumber \\
   \frac{d\Psi(t)}{dt} &=& \alpha_{\mathrm{off}} \left[1 - \Psi(t) \right]\cdot (1 - u(t)) \nonumber \\ 
   & & \quad + \; \alpha_{\mathrm{on}} \left[K_{\Psi}(t) \Psi_{\mathrm{inf}} - \Psi(t) \right]\cdot u(t), \nonumber 
\end{eqnarray}
where $K_{\Psi}$ is a time-varying gain given by 
\begin{equation*}
    K_{\Psi}(t) = 1 - \gamma_k \gamma_A \frac{\rho}{1-\rho} \frac{I(t)}{N} 
\end{equation*}  
In~(\ref{eq:nonlinear-eqns}), $N$ is the population which can be reasonably assumed to be constant, $\beta$ is the average number of contacts that is sufficient for viral transmission per day, $\gamma_A$ and $\gamma_B$ are the recovery rates from infected and hospitalisation respectively, $\lambda$ is the hospitalisation rate and $\rho$ is the fatality rate.  These parameters are assumed to be constant for the duration of the wave. The settling time $\alpha_{\mathrm{on}}$ and $\alpha_{\mathrm{off}}$ relate the response time of the population to the isolation measures imposed. The parameter  $\Psi_{\mathrm{inf}}$ reflects the ``strongest'' level of practical isolation with $K_\Psi$ being a time-varying isolation gain that is dependent on the number of infections while $\gamma_k > 0$ is used to parameterise this relationship. 

The control signal $u$ determined by the MPC and adopted by the authorities is implemented through a range of measures. To decide on the social activities allowed for each of the isolation levels, we refer the reader to \cite{Cascini2022} wherein a ``containment index'' is used. In \cite{Cascini2022}, nine different activity areas are identified which included schools, business and workplace, shops, hospitality, personal care, internal movement among others and a score between $0$ and $1$ is given to each of the areas in terms of severity of restriction imposed to assess their overall impact on the propagation of the virus. The total restriction level nationwide / statewide, referred to as the ``containment index'' is then taken as the average of the restrictions imposed in each of these areas. The idea behind the containment index originates from the Oxford COVID-19 Government Response Tracker study which documents the effects of government policies related to closure and containment in 180 countries and correlates the timing of policy adoption, policy easing and re-imposition on epidemiological indicators \cite{Hale2021}.  

\section{Methodology} 
\label{sec:methodology}

\subsection{Nonlinear System Identification} \label{sec:NSI}

The parameters of the nonlinear model are determined using genetic algorithm from data gathered when Mauritius was in a state of partial lockdown. The mathematical framework behind genetic algorithms was developed and first presented in \cite{Holland1975}. Since then, they have been used in diverse areas including system identification as early as in the 1980s \cite{Das1988, Kristinsson1992} and many more recently, e.g., see \cite{Khandelwal2019, Jain2024}. In our case, the genetic algorithm begins by creating a random initial population of parameterisations of model~(\ref{eq:nonlinear-eqns}). A fitness value is then computed for each individual $p$ in the population reflecting how good this individual is in representing the underlying dynamics. Individuals from the current population with the highest fitness values (elites) are selected to become the new parents which then produce the children for the next generation. 

To compute the fitness of an individual, we use the weighted sum of the Euclidean distance between the data points $x_{\mathrm{data}}$ and the fitted curve $f(t,p,x_0)$ generated by the individual $p$ in the current population at time $t$ starting from the initial condition $x_0$. To this end, we use the data available from \cite{WHO2024} and \cite{moh2022} comprising the number of new infected cases (daily), the number of new hospitalised cases (weekly) and the total number of deaths (daily). We first interpolate between successive weekly data points for the hospitalised cases to generate daily data points and then calculate the corresponding state in the Infected and Hospitalised compartments using the mean infectious time and the mean duration of hospital stay after admission. They are assumed to be 10 days (see~\cite{Byrne2020,Johansson2021}) and 15 days (see~\cite{Alimohamadi2022}) respectively. We then find the optimal individual $p$ by minimising the following function:
\begin{equation}
    \min_p \sum_{\substack{x_{data} \in \mathcal{D}\\ s = i, h, d}} \alpha_s || (f_s(t, p, x_0) - x_{\mathrm{data},s})||_2 \label{eq:min-nsi}
\end{equation}
where $\mathcal{D}$ is the set of all available data points and $\alpha_i$, $\alpha_h$ and $\alpha_d$ are the weights associated with the set of infected, hospitalised and dead data respectively. The process to obtain the best solution over successive generations is presented in Algorithm~\ref{tab:ga}.

\begin{algorithm}[]
\caption{\textsc{Genetic Algorithm}}
\label{tab:ga}
\begin{algorithmic}[]
\STATE \textbf{Initialize Population} Generate an initial population of random genomes with genes $[\beta, \gamma_A, \gamma_B, \rho, \lambda, \alpha_{\mathrm{on}}, \Psi_{\mathrm{inf}}, \gamma_k]$.
\STATE \textbf{Evaluate Population} Calculate the fitness value of each genome in the population
\begin{equation*}
\sum_{\substack{x_{data} \in \mathcal{D}\\ s = i, h, d}} \alpha_s || (f_s(t, p, x_0) - x_{\mathrm{data},s})||_2
\end{equation*}
\STATE \textbf{Repeat} until stopping condition is met
       \begin{enumerate}
        \item \textit{Selection} : Select parent solution with the highest fitness values from the population.
        \item \textit{Crossover} : For each pair of parents, perform crossover (recombination) to create offspring with a given crossover probability,
        \item \textit{Mutation} : For each offspring, apply random changes with a given mutation probability to maintain genetic diversity.
        \item \textit{Evaluate New Population} : Calculate the fitness value of each genome in the new population. 
       \end{enumerate}
\STATE \textbf{Return} the best genome after convergence of solution to a tolerance of $1\times 10^{-6}$ or after 300 generations.
\end{algorithmic}
\end{algorithm}

\subsection{Model Predictive Control} \label{sec:MPC}
We develop an MPC scheme using the identified model obtained in Section~\ref{sec:NSI}. MPC is a constrained optimal control strategy that computes the sequence of optimal control inputs that needs to be applied to a system at each iteration step over a finite prediction horizon by solving a constrained optimization problem. In this way, the behaviour of the system can be accurately controlled within the prediction horizon. The mathematical model of the system as well as the current state measurements (or estimates) are fed to the algorithm to predict and optimize the future behavior of the system. At each time instant $k$ , we consider the optimization problem over the next $N$ instants where $N$ is the prediction horizon, i.e., we compute the sequence of control inputs over the prediction horizon but only the first element of the sequence of inputs is applied to the system. This process is repeated over the duration of the simulation $N_T$; see e.g. \cite{Kouvaritakis2016}. MPC is particularly well suited to the problem at hand since it can systematically consider the optimization requirements (e.g. minimizing the number of infected/dead people) while satisfying constraints (e.g. maximum number of hospitalized people) and generate restricted control signals.

Using the nonlinear continuous time SIHRDC model identified in Section~\ref{sec:NSI} within the MPC framework would lead to a nonconvex optimization problem. To preserve the convexity of the resulting MPC problem, we linearize and discretize the SIHRDC model at each time instant around the current state. The control signal determines the social isolation policy to regulate the level of human interaction in the community. Let the discrete-time linear time-invariant (LTI) system be represented as:
\begin{equation}
    x_{k+1} = A_k x_k + B_k u_k,
\end{equation}
where $k \in \{ 0, 1, \cdots, N \}$ is the discrete time instant, $x_k \in \mathbb{R}^6$  is the state vector comprising of the states $S, I, H, R, D$ and $\Psi$, $u_k \in \mathbb{Z}$ is the control input reflecting the isolation level, and $A_k \in \mathbb{R}^{n \times n}$ and $B_k \in \mathbb{R}^{n \times m}$ are the state and input matrix respectively at time instant $k$ computed by linearizing the model around the current state and discretizing it with a sampling time of 1 day. 

Our cost function penalises the number of infected people $I$, the number of hospitalised people $H$ and the number of new fatalities $D$ over the prediction horizon. We also penalise the control input $u$ to minimise social and economic repercussions. To this end, we choose the cost function as: 
$$
J = \sum_{k=0}^{N-1} q_1 I_k^2 + q_2 H_k^2 + q_3 (D_k - D_{k-1}) + r u_k^2
$$
where $q_1, q_2, q_3$ and $r$ are positive scalars. Note that we consider here the difference in the number of casualties since $D_k$ is the total number of dead individuals until time $k$ unlike $I_k$ and $H_k$ which are the number of infected and hospitalized individuals at time $k$. The cost function can also be written in the more general form 
\begin{equation}
    J = \sum_{k=0}^{N - 1} x_k^T Q_a x_k + \sum_{k=1}^{N} (x_k - x_{k-1})^T Q_b (x_k - x_{k-1}) + u_k^T R u_k \label{eq:penalty}
\end{equation}
where $Q_a = \operatorname{diag}(0,q_1,q_2,0,0), \, Q_b = \operatorname{diag}(0,0,0,0,q_3)$ and $R = r$. For a system with initial state $x_i$, we determine the optimal sequence of control inputs by solving the online optimal control problem:
\begin{eqnarray}
    \min_{u_0, u_1, \cdots, u_{N-1}} & & J(x_0,\hdots,x_{N+1},u_0,\hdots,u_N) \nonumber \\
    \mathrm{subject\;to} \; \quad & & x_0 = x_i, \ x_N \in \mathbb{X}, \nonumber \\
    & &  x_{k+1} = A_k x_k + B_k u_k, \label{eq:optimization-formulation} \\
    & & x_k \in \mathbb{X}, \ u_k \in \mathbb{U}, \nonumber \\
    & & \text{for all } k \in \{0,\hdots,N-1\} \nonumber
\end{eqnarray}
For simplicity, the set $\mathbb{U}$ restricts the control signal $u_k$ to have specific values of $1/(n-1)$ between $0$ and $1$, where $n$ represents the number of level of isolation policies adopted by the authorities while the set $\mathbb{X}$ applies an upper bound on the number of hospitalized individuals depending on the capacity of the hospitals.

\section{Results} \label{sec:Results}
\subsection{Nonlinear System Identification}
We identify the parameters of the nonlinear model~(\ref{eq:nonlinear-eqns}) using Algorithm~\ref{tab:ga}. The parameter set is $ p = [\beta, \gamma_A, \gamma_B, \rho, \lambda, \alpha_{\mathrm{on}}, \Psi_{\mathrm{inf}}, \gamma_k]$ and as such each individual in the population considered in Algorithm~\ref{tab:ga} has a size of 8. We restricted the number of new population to a maximum of $300$. The population size $a$ was varied between $80$ to $120$ in incremental steps of $5$. Furthermore, due to the randomness of genetic algorithm, we repeat the parameter estimation three times for each population size. In determining the parameters, more emphasis was laid on matching the number of fatalities which was accurately known. The number of infections is subjected to sub-notifications due to the fact that authorities are only officially aware of infectious cases for which tests have been conducted in designated health centres. Hence, we used $\alpha_1 = 1 $ for the infected cases, $\alpha_2 = 1 $ for hospitalised cases and a larger weight $\alpha_3 = 2 $ for the number of deaths. The parameters determined are given in Table~\ref{tab:parameters-identified}.
\begin{table}[htb]
  \centering
    \begin{tabular}{||c|c|c|c||} \hline 
    \multirow{2}{4.5em}{Model Parameters} & \multirow{2}{3.5em}{Values} & \multirow{2}{4.5em}{Control Parameters} & \multirow{2}{3.5em}{Values} \\[10pt] \hline \hline
      $\beta$ & 0.63651 & $\alpha_{\mathrm{on}}$ & 1.9498 \\ \hline
      $\gamma_A$ & 0.40575 & $\alpha_{\mathrm{off}}$ & 3.8996  \\ \hline
      $\gamma_B$ & 0.09333 & $\Psi_{\inf}$ &  0.60376 \\ \hline
      $\rho$ & 0.07745 & $\gamma_k$ & 21.632 \\ \hline
      $\lambda $ & 0.00299 & &     \\ \hline
  \end{tabular}
  \caption{Identified Parameters}
  \label{tab:parameters-identified}
\end{table}

\begin{figure}[htb]
  \begin{center}
    \includegraphics[scale = 0.65]{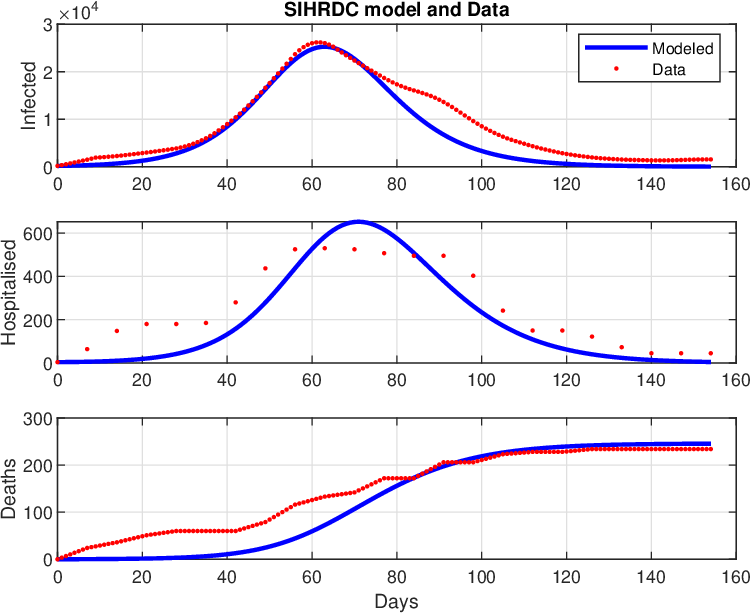}
    \caption{Identified Model and Actual Data}
    \label{fig:third-wave-model-data}
  \end{center}
\end{figure}

Figure~\ref{fig:third-wave-model-data} shows a plot of the response of the identified model with the corresponding data points. It should be noted that this model was identified assuming $u_k=0.5$, corresponding to restriction measures that were maintained for the entire duration of the third wave.

\begin{figure}[htb]
  \begin{center}
    \includegraphics[scale = 0.65]{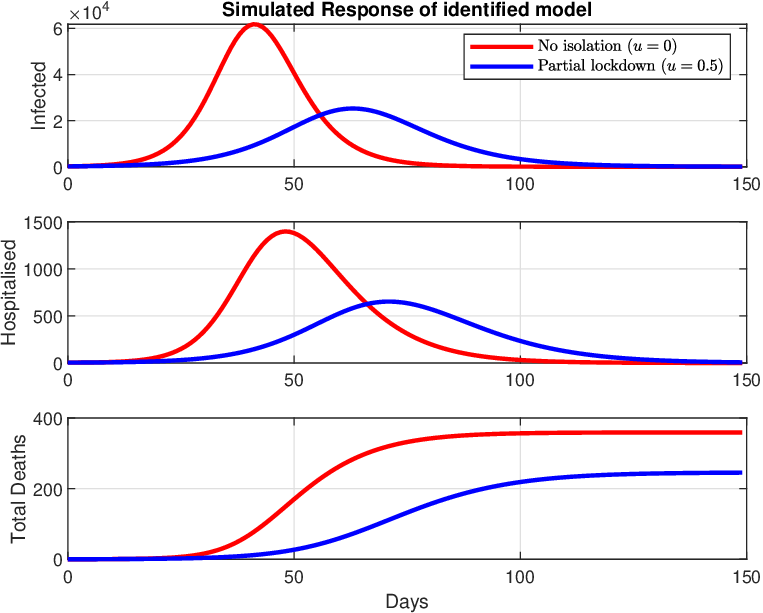}
    \caption{Under partial lockdown ($u=0.5$) and without any isolation measures ($u=0$).}
    \label{fig:comparison_u0_uhalf}
  \end{center}
\end{figure}

The open-loop response of the model ($u_k=0$) would present the worst-case scenario. In Figure~\ref{fig:comparison_u0_uhalf}, this scenario is compared to the case where partial isolation was in place ($u_k=0.5$). Without any isolation measures, the number of infected individuals would shoot up to nearly 70,000 cases, requiring facilities to accommodate a peak of 1500 hospitalisations, and the resulting number of deaths settling to 350 at the end of the wave. It should also be noted that the peak values of infection/hospitalisation occurs quite swiftly (within 30 days for $u_k=0$), implying a large burden on health centres. 

\subsection{Controller design}
The spread of SARS-CoV-2 is governed by a relatively ``slow'' dynamics with measurements of new infections and deaths available on a daily basis. As such, the continuous-time system is discretised with a sampling time of $T_s =$ 1 day using Euler's method. We simulate the system for the whole duration of the third wave; i.e., $N_T = 150$ days. Since new isolation policies need to be maintained for a minimal number of days, we also require that any change in $u_k$ remain valid for a minimum of 7 days. Hence, we choose the control horizon $N$ to be 14 days which allows sufficient error margin for decision making on a weekly basis. Furthermore, we also limit the maximum number of beds to 200 (i.e., $x_{3} \leq 200$ in (\ref{eq:optimization-formulation})). 

\begin{figure}[htb]
  \begin{center}
    \includegraphics[scale = 0.6]{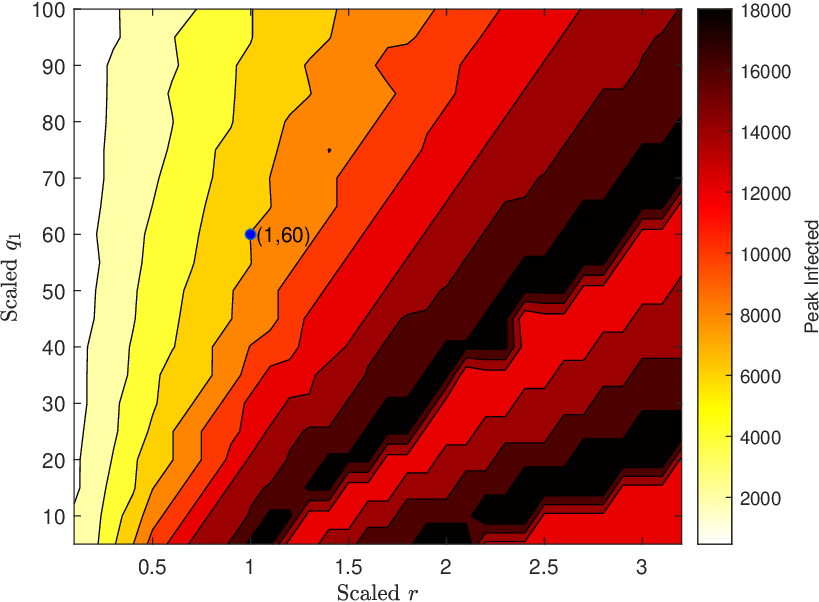}
    \caption{Effect of varying $q_{1}$ and $R$ on peak number of infections.}
    \label{fig:max-infections}
  \end{center}
\end{figure}

To determine a good choice of weights to reduce casualty while allowing for increased social interaction and economic activities, we vary the weighting matrices $Q_a$, $Q_b$ and $R$ in~(\ref{eq:penalty}). The parameters $q_1$ and $q_2$ are used to penalise the number of infections and hospitalisations respectively. We find that the parameter $q_2$ has minimal effect on the solution once $q_1$ is fixed. Similarly, the parameter $q_3$ used to minimize daily casualties has little impact on the solution once $q_1$ is fixed, which is expected since the model assumes that casualties only occurred after hospitalisation. As such, we vary $q_1$ and $R$ within suitable ranges over which the optimisation problem has feasible solutions and quantify the effect of the weights on the maximum number of infections and total number of deaths. The weights $q_1$, $q_2$, and $q_3$ are scaled by the factors $1/(26000)^2, 1/(525)^2$ and $1/(240)^2$ respectively, corresponding to the peak identified values of $I, H$ and $D$. 
Figure~\ref{fig:max-infections} confirms that decreasing $q_1$ and increasing $R$, which corresponds to reducing the penalty due to the number of infected people and minimising the extent and/or duration of isolation periods respectively, leads to greater number of infections. 

\begin{figure}[htb]
  \begin{center}
    \includegraphics[scale = 0.65]{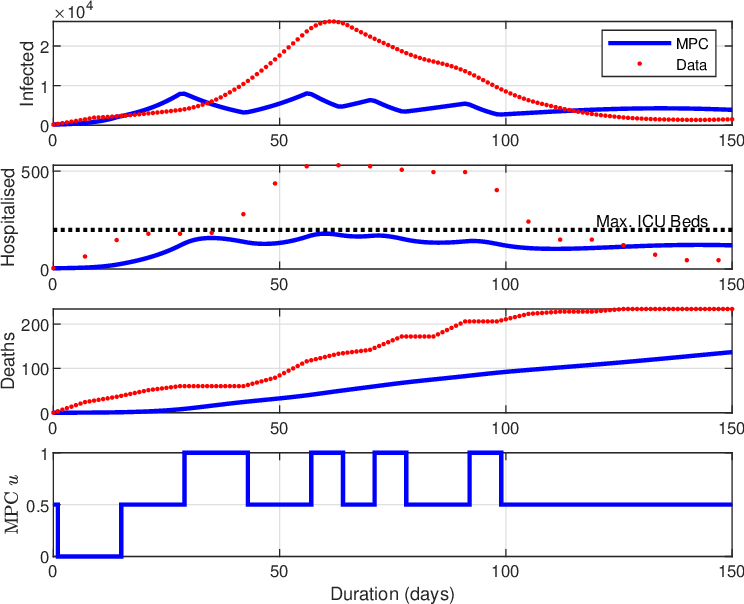}
    \caption{Comparison of epidemic response with proposed MPC and actual data.}
    \label{fig:mpc-control-isolation}
  \end{center}
\end{figure}

We selected the scaled design parameters $q_1 = 60$ and $R = 1$ from Figure~\ref{fig:max-infections} and simulated the response of the system keeping $q_2 = q_3 =  1$. Figure~\ref{fig:mpc-control-isolation} shows the improvement achieved with the MPC in place. With short variations of the isolation level at the appropriate time, the maximum infections were reduced from 26000 to 9000, and the maximum hospitalisations were reduced from 525 to 190. The number of casualties also decreased to 140 at the end of the 150 day period. Interestingly, this result was achieved with $u$ switching to zero thus removing all isolation measures at the beginning of the epidemic wave. Further improvement can be easily obtained by manipulating the weights $q_1$ and/or $R$, but it will be at the expense of longer and more frequent periods of stricter social isolation. The aim here is to show that with minimal additional economic and social burden, the epidemic can be better controlled with about 40\% reduction in the number of fatalities while allowing health centres to operate within the limit of their capacity.

\begin{figure}[htb]
  \begin{center}
    \includegraphics[scale = 0.6]{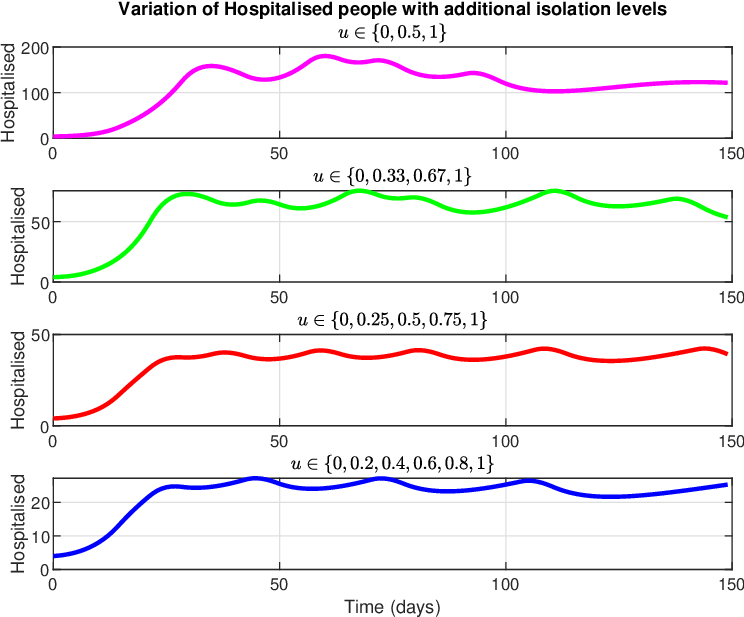}
    \caption{Performance of controller with varying levels of social isolation.}
    \label{fig:perf-levels}
  \end{center}
\end{figure}

Furthermore, having additional levels of isolation policies to control the spread of the virus is expected to improve performance. As before, we selected $q_1 = 60$ and $R = 1$ and simulated the system for 150 days. Figure~\ref{fig:perf-levels} shows the effect of including additional isolation policies on the number of hospitalisations. The risk of exceeding the 200-bed threshold is more pronounced with only 3 isolation levels $\{0, 0.5, 1\}$ in use. Much better performance is achieved with 4 or more isolation levels used with considerably less hospitalisations, making it a good case for authorities to justify the administrative burden of imposing more layers of control through the use of additional containment index~\cite{Cascini2022}.

\section{Conclusion and Future Work}
\label{sec:Conclusion}
A data-driven MPC scheme has been proposed to better mitigate the impact of a future pandemic, with a case study built using statistical data gathered during the most prominent COVID-19 wave in Mauritius. A nonlinear dynamical system representing the propagation of the virus as well as the population response was parameterised using genetic algorithms. Our approach shows that the hospital occupancy level remained below the threshold level of 200 by adopting isolation levels specified by the MPC, rather than a fixed isolation level with the partial lockdown adopted by the authorities. In turn, the number of casualties is reduced from 240 to 140. Furthermore, we investigated the effect of introducing additional containment levels and show that there is room for the epidemic to be better managed even further albeit with additional administrative burden on the authorities. One avenue for future work will involve addressing the limitation of having a grey-box model at the onset with a Gaussian process model that is updated as new data is gathered.


\printbibliography
\end{document}